\documentclass{jfm}
\usepackage{graphicx}
\usepackage{epstopdf, epsfig}
\usepackage{xparse}

\usepackage{amsmath}
\usepackage{amsfonts}
\usepackage[version=4]{mhchem}
\usepackage{siunitx}
\usepackage{longtable,tabularx}
\usepackage{physics}
\usepackage{mathrsfs}
\usepackage{gensymb}
\usepackage{natbib}

\usepackage{tabularx}
\usepackage{multirow}
\usepackage{hyperref}
\usepackage[noabbrev,nameinlink]{cleveref}
\hypersetup{
    colorlinks = true,
    citecolor=blue,
    linkcolor = blue
    }
\usepackage{float}

\usepackage{mathtools}
\usepackage{caption}
\usepackage{subcaption}
\usepackage{tikz,siunitx,mwe}

\crefformat{equation}{(#2#1#3)}

\usepackage{amssymb}
\usepackage{booktabs}
\usepackage{bbm}

\newcommand{\scalebarimg}[3]{
  \begin{tikzpicture}
    \draw node[name=micrograph] {\includegraphics[width=#2\textwidth]{#1}}; 
    \draw  (micrograph.north west)  node[anchor=north west,xshift=-3,yshift=-1,black]{(\textit{\small{#3}})}; 
  \end{tikzpicture}
}

\newcommand{\scalebarimgtr}[3]{
  \begin{tikzpicture}
    \draw node[name=micrograph] {\includegraphics[trim={0 2cm 0 0},clip,width=#2\textwidth]{#1}}; 
    \draw  (micrograph.north west)  node[anchor=north west,xshift=-3,yshift=-1,black]{(\textit{\small{#3}})}; 
  \end{tikzpicture}

}

\usepackage{algorithm,amsmath,algpseudocode,algcompatible,setspace,lmodern,pdftexcmds}

\DeclareMathAlphabet{\mathsfbi}{OT1}{\sfdefault}{bx}{sl}
\DeclareMathVersion{sfletters}
\SetSymbolFont{letters}{sfletters}{OML}{ntxsfmi}{b}{it}

\makeatletter
\newcommand{\mathbfsbilow}[1]{%
  \text{\mathversion{sfletters}$\m@th#1$}%
}

\DeclareRobustCommand{\tensor}[1]{%
  \begingroup
  \ifcat\noexpand #1\relax
    \edef\greek@test{\detokenize{#1}}%
    \edef\greek@test{\expandafter\@cdr\greek@test\@nil}%
    \edef\greek@test{\expandafter\@car\greek@test\@nil}%
    \edef\x{\the\lccode\expandafter`\greek@test}%
    \edef\y{\number\expandafter`\greek@test}%
    \ifnum\x=\y\relax
      \mathbfsbilow{#1}%
    \else
      \mathsfbi{#1}%
    \fi
  \else
    \mathsfbi{#1}%
  \fi
  \endgroup
}
\makeatother

\usepackage{etoolbox}

\makeatletter

\patchcmd{\NAT@citex}
  {\@citea\NAT@hyper@{%
     \NAT@nmfmt{\NAT@nm}%
     \hyper@natlinkbreak{\NAT@aysep\NAT@spacechar}{\@citeb\@extra@b@citeb}%
     \NAT@date}}
  {\@citea\NAT@nmfmt{\NAT@nm}%
   \NAT@aysep\NAT@spacechar\NAT@hyper@{\NAT@date}}{}{}

\patchcmd{\NAT@citex}
  {\@citea\NAT@hyper@{%
     \NAT@nmfmt{\NAT@nm}%
     \hyper@natlinkbreak{\NAT@spacechar\NAT@@open\if*#1*\else#1\NAT@spacechar\fi}%
       {\@citeb\@extra@b@citeb}%
     \NAT@date}}
  {\@citea\NAT@nmfmt{\NAT@nm}%
   \NAT@spacechar\NAT@@open\if*#1*\else#1\NAT@spacechar\fi\NAT@hyper@{\NAT@date}}
  {}{}

\makeatother

\shorttitle{Optimal global natural receptivity}
\shortauthor{O. Kamal et al.}

\title{Global receptivity analysis: physically realizable input-output analysis}

\author{Omar Kamal\aff{1}
  \corresp{\email{okamal@caltech.edu}}, Matthew T. Lakebrink\aff{2}
 \and Tim Colonius\aff{1}}

\affiliation{\aff{1}Department of Mechanical and Civil Engineering, California Institute of Technology,
Pasadena, CA 91125, USA
\aff{2}The Boeing Company, Hazelwood, MO 63042, USA}

\begin{document}

\maketitle

\begin{abstract}
In the context of transition analysis, linear input-output analysis determines worst-case disturbances to a laminar base flow based on a generic right-hand-side volumetric/boundary forcing term. The worst-case forcing is not physically realizable, and, to our knowledge, a generic framework for posing physically-realizable worst-case disturbance problems is lacking. In natural receptivity analysis, disturbances are forced by matching (typically local) solutions within the boundary layer to outer solutions consisting of free-stream vortical, entropic, and acoustic disturbances. We pose a scattering formalism to restrict the input forcing to a set of realizable disturbances associated with plane-wave solutions of the outer problem. The formulation is validated by comparing with direct numerical simulations (DNS) for a Mach 4.5 flat-plate boundary layer. We show that the method provides insight into transition mechanisms by identifying those linear combinations of plane-wave disturbances that maximize energy amplification over a range of frequencies. We also discuss how the framework can be extended to accommodate scattering from shocks and in shock layers for supersonic flow. 
\end{abstract}


\vspace*{-1.2cm}

\section{Introduction}

Boundary-layer receptivity analyses determine how free-stream vortical, entropic, and acoustic waves excite instabilities. Several approaches have been developed to accomplish this inner-to-outer matching, such as forcing a flat-plate boundary layer with an induced traveling wave having a frequency of an incident acoustic wave and a wavenumber associated with surface irregularities \citep{crouch_1992}. However, many studies \citep{goldstein_1983,duck_1996,qin_2016,ruban_2021} still rely upon asymptotic expansions, which often require additional approximations such as restrictions to low frequencies \citep{fedorov_2003}. Although DNS can alleviate these challenges, many (expensive) calculations are needed to characterize the dominant natural receptivity mechanisms. This is especially apparent for design problems where the inverse study is often more useful: what are the worst-case disturbances that lead to maximal amplification?

{\em Input-output} analysis provides a framework for this kind of optimization problem by determining surface or volumetric inhomogeneities, i.e. inputs, that lead to maximal amplification of disturbances, i.e. outputs. \citet{Trefethen_1993} introduced studying the pseudospectra of the ``linearized Navier-Stokes evolution" operator as a tool for understanding non-modal amplification of disturbances in Couette and Poiseuille flows. \citet{monokrousos2010global} used the input-output framework to determine optimal amplification in the spatially-evolving flat-plate boundary layer, and it has subsequently been used in a variety of contexts, including extensions to computing optimal disturbances in turbulent mean flows \citep{schmidt_2018}. The framework has also been adapted to high-speed compressible flows \citep{nichols2011global,nichols2019input,cook2020matrix,cook_nichols_2022,lugrin_2021,bugeat20193d,bae_2020}. Furthermore, there have been contemporary methodological advancements pertaining to the nature of the optimal forcing, such as using sparsity-promoting norms in computing localized forcing structures \citep{skene_2022}. 

However, while the inputs can be readily restricted to subspaces by, for example, forcing only in certain equations (mass, momentum, or energy), and/or in certain flow regions (e.g. \citet{jeun_2016}), the resulting inhomogeneous problem is not {\it physically realizable}, in the sense that the sources are unconnected to any physical mechanism that produces them. In this work, we employ a scattering formalism to restrict input-output analysis to forcings that are associated with free-stream disturbances. We decompose the full linear solution into an incident component, representing vortical, entropic, or acoustic disturbances to the free-stream, and a scattered (or residual) component that is forced by the incident wave propagated through the linearized equations. This forcing approaches zero in the free-stream where the incident waves satisfy the governing equations, but is nonzero within the shock- and boundary-layer regions where it can be parameterized and optimized using the standard input-output (singular value decomposition) framework. This permits natural receptivity analysis to be performed directly in the global framework without recourse to asymptotic analysis (though with its own challenges as we discuss).

The global optimal receptivity formalism is developed in \S~\ref{sec:method}, after which the problem is simplified to high-speed flow over a 2D flat plate (ignoring any shock) in order to validate the methodology with previous results. In \S~\ref{sec:in_out_fast}, we employ the input-output scattering formalism with various free-stream waves and analyze the corresponding responses. Thereafter, we compute the optimal incident waves that maximize 2D disturbance-energy amplification for a Mach 4.5 flat-plate boundary layer over a range of frequencies, showing how the input-output framework complements and extends traditional receptivity theory. In \S~\ref{sec:discussion}, we summarize and discuss future work.   

\section{Methodology}\label{sec:method}

\subsection{Scattering ansatz}\label{sec:scattering}

We begin with the non-dimensional, fully compressible Navier-Stokes equations, linearized about a time-independent base flow such that
\begin{equation}
    \boldsymbol {q}(x,y,z,t)=\boldsymbol{\bar{q}}(x,y,z)+\boldsymbol{q'}(x,y,z,t),
\end{equation}
where $\boldsymbol {q}=[\rho,u,v,w,T]^T$ is the state vector and $(x,y,z)$ is the streamwise, wall-normal, and spanwise directions, respectively. We take these variables as non-dimensionalized by the free-stream density $\rho^*_\infty$, sound speed $c^*_\infty$, and temperature $c^{*2}_\infty/{c^*_p}_\infty$, where $^*$ represents dimensional quantities. Note that length scales are normalized with $\delta^*_0= \sqrt{\left( {\nu^*_\infty x^*_0} / {U^*_\infty} \right)}$, where $x^*_0$ is the inlet $x$-coordinate.

After analytically linearizing about the steady laminar flow, we have
\begin{equation}
\begin{aligned}
    {\mathsfbi G}\frac{\partial \boldsymbol{q'}}{\partial t} + {\mathsfbi {A_x}}\frac{\partial \boldsymbol{q'}}{\partial x} +  {\mathsfbi {A_y}}\frac{\partial \boldsymbol{q'}}{\partial y} +& {\mathsfbi {A_z}}\frac{\partial \boldsymbol{q'}}{\partial z} +{\mathsfbi A}\boldsymbol{q'} +{\mathsfbi {A_{xx}}}\frac{\partial^2 \boldsymbol{q'}}{\partial x^2} +{\mathsfbi {A_{yy}}}\frac{\partial^2 \boldsymbol{q'}}{\partial y^2} +{\mathsfbi {A_{zz}}}\frac{\partial^2 \boldsymbol{q'}}{\partial z^2} + \\ & {\mathsfbi {A_{xy}}}\frac{\partial^2 \boldsymbol{q'}}{\partial x \partial y} + {\mathsfbi {A_{xz}}}\frac{\partial^2 \boldsymbol{q'}}{\partial x \partial z} + {\mathsfbi {A_{yz}}}\frac{\partial^2 \boldsymbol{q'}}{\partial y \partial z} = 0.
\end{aligned}
\end{equation}
We next enter the stationary frequency domain via the Laplace transform with $s=-\mathrm{i}\omega$, and after globally discretizing the equations and incorporating appropriate boundary conditions, the resulting equations may be written in matrix form
\begin{equation} \label{eq:LNS}
{\mathsfbi L} {\hat {\boldsymbol q}} =0,
\end{equation}
where $\boldsymbol{\hat q}$ is the discretized perturbed state vector, ${\mathsfbi L}$ is the linearized Navier-Stokes (LNS) matrix, and the hat indicates the frequency domain. For now, we imagine that the domain is the entire region external to a given immersed body, and is endowed with no-slip at the surface and disturbances decaying to zero at infinity, where the base flow becomes a uniform flow (in the $x$-direction).  Practical boundary condition issues are discussed later.

\begin{figure}
  \centering
  \begin{subfigure}[H]{0.35\textwidth}
     \centering
{\phantomsubcaption\label{fig:schm1}
    \scalebarimg{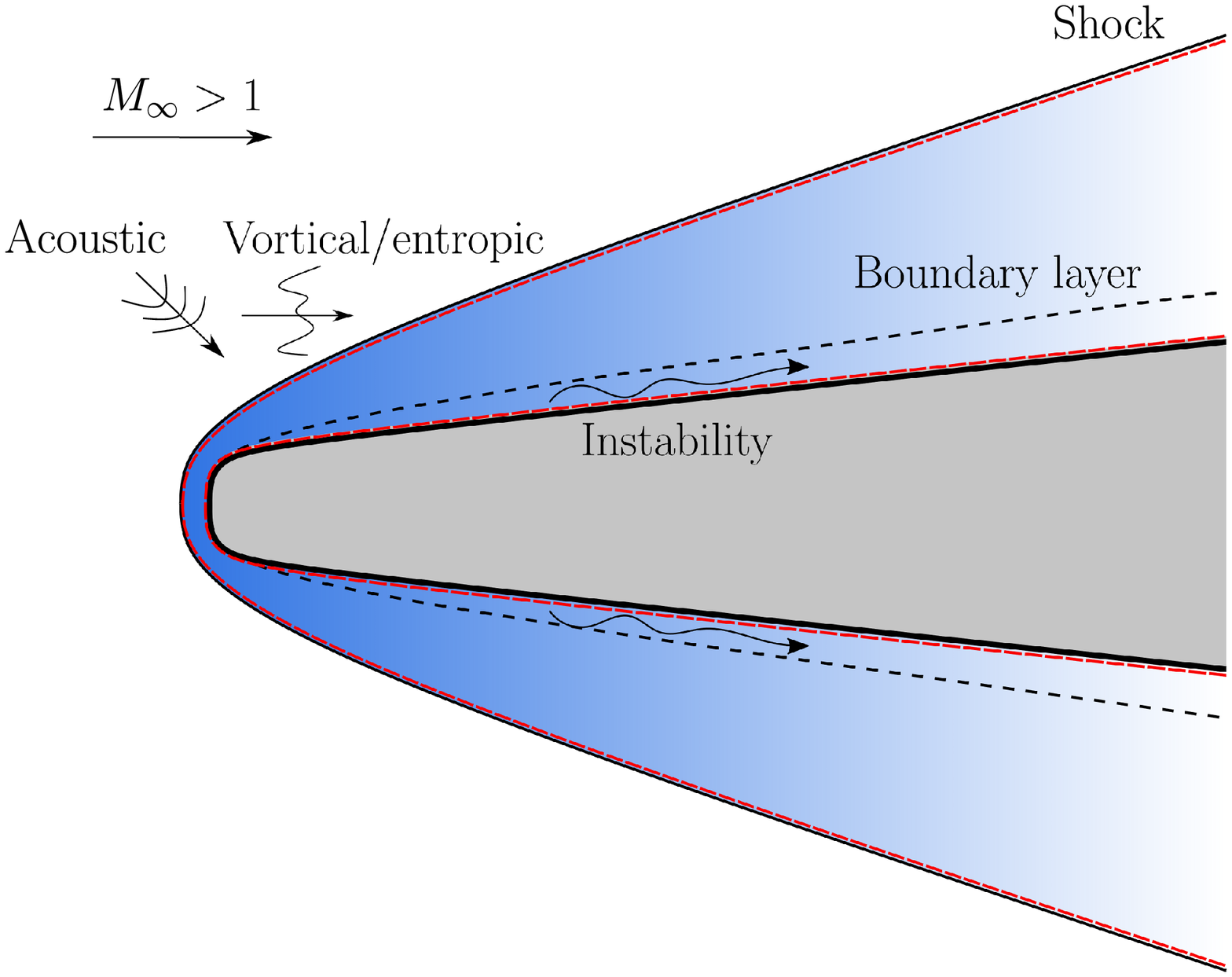}{1}{a}}
  \end{subfigure}
  \hspace*{1cm}
  \begin{subfigure}[H]{0.35\textwidth}
     \centering
    {\phantomsubcaption\label{fig:schm2}
    \scalebarimg{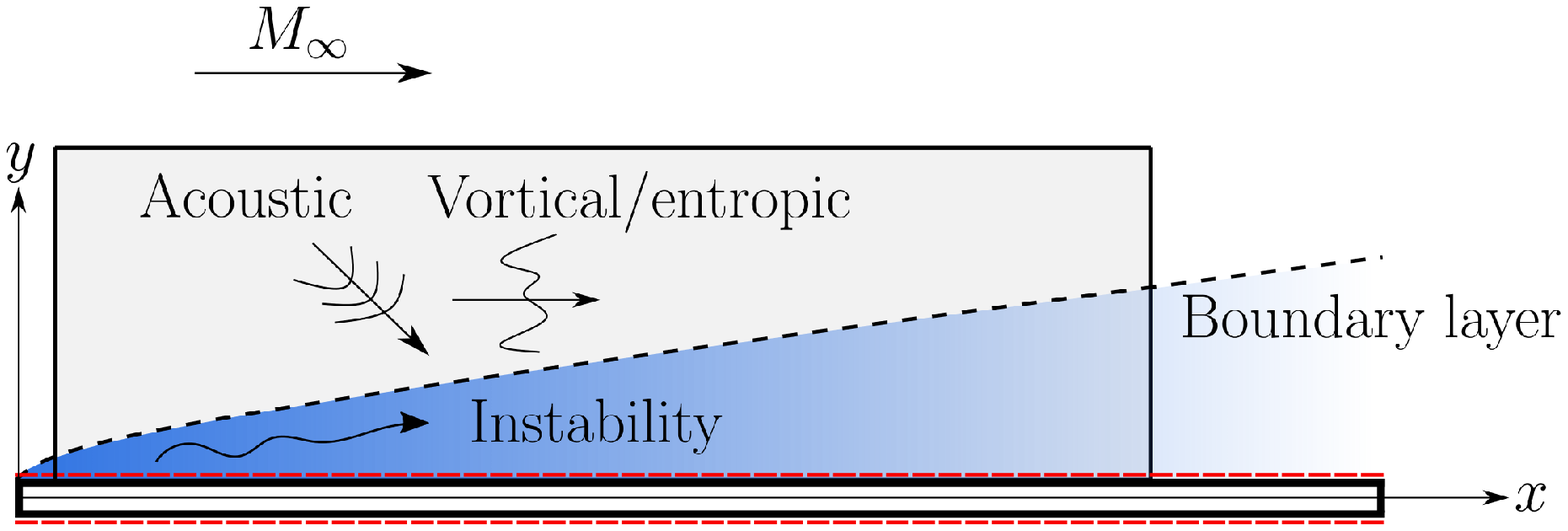}{1}{b}}
 \end{subfigure}
 \vspace{-2\baselineskip}
\caption{A depiction of $\mbox{supp}( {\mathsfbi L} \boldsymbol{\hat q}^{\boldsymbol i} )$;  (\textit{a}) supersonic case generally; (\textit{b}) idealized supersonic flat plate without shock layer. Depth of the blue shaded region corresponds to the strength of $\text{supp}({\mathsfbi L} \boldsymbol{\hat q}^{\boldsymbol i})$, whereas the red dashed lines indicate surface scattering from the body and shock. The grey shaded region in (\textit{b}) corresponds to the computational domain utilized.}
\label{fig:schematic}
\end{figure}

We wish to study how incident acoustic, vortical, or entropic waves scatter into instability waves due to the presence of the immersed surface and associated shock/boundary layer. Without loss of generality, we can decompose the solution into incident and scattered components, ${\boldsymbol{\hat q}} = \boldsymbol{\hat q}^{\boldsymbol i} + \boldsymbol{\hat q}^{\boldsymbol s}$, so that
\begin{equation}\label{eq:in)out}
 {\mathsfbi L} \boldsymbol{\hat q}^{\boldsymbol s}  = - {\mathsfbi L} \boldsymbol{\hat q}^{\boldsymbol i} \equiv \boldsymbol {\hat f}.
\end{equation}
If the incident component is known, then the scattered component can be determined. We thus further specify that the incident component takes the form of appropriate linear vortical, entropic, or acoustic waves in a uniform flow, such that ${\mathsfbi L} \boldsymbol{\hat q}^{\boldsymbol i} \rightarrow 0$ in the free-stream. The support of the forcing is shown in \cref{fig:schematic}, and is confined to the shock- and boundary-layer regions for supersonic flow, depicted by volumetric sources (blue) and surface sources (red).  In the discretized case, these are not distinct and are both incorporated directly in ${\mathsfbi L} \boldsymbol{\hat q}^{\boldsymbol i}$. The source originating at the shock surface includes the reflection and transmission of incident disturbances of each type to every other.  In the linearized framework proposed here, we are implicitly linearizing about a fixed shock position, and this would also neglect second-order effects associated with shock oscillations  \citep{cook_nichols_2022}.

The shock also gives rise to technical challenges since we are discretizing about a discontinuous solution. As a first step towards establishing the general framework, in what follows, we limit further analysis to the flat-plate scenario shown in the \cref{fig:schematic}(\textit{b}), where any shock and shock layer are neglected and ${\mathsfbi L} \boldsymbol{\hat q}^{\boldsymbol i}$ decays smoothly towards infinity (similar to the scenario in subsonic flow).  We choose the computational domain depicted in the sketch, which also neglects scattering sources from any leading-edge geometry, and scattered waves that are generated from below and diffracted around the plate, which are expected to be small compared to direct irradiation.  We choose the downstream extent of the computational domain on physical grounds so that dominant instability mechanisms (as a function of Reynolds number) are captured within the domain. %Lastly, any artificially-induced scattering at the far-field due to the imposition of boundary conditions in ${\mathsfbi L}$ is neglected (these boundary conditions, such as homogeneous Dirichlet conditions, are usually in conflict with the incident disturbances).

Three potential sources of error may be identified in our framework. The first of these is discretization error, which is controlled by choosing a sufficiently fine grid.  However, typical grids that are desirable for the scattered-field solution, i.e. ones that are highly stretched outside the boundary layer, present a challenge for incident waves of sufficiently high frequency, in that a direct computation of ${\mathsfbi L} \boldsymbol{\hat q}^{\boldsymbol i}$ in this region is prone to large errors.  This is alleviated by computing ${\mathsfbi L} \boldsymbol{\hat q}^{\boldsymbol i}$ on a much finer grid and then interpolating the results onto the coarser computational mesh used for the solution of the scattered field. The remaining two errors are associated with posing the correct outer solution for ${\mathsfbi L} \boldsymbol{\hat q}^{\boldsymbol i}$. It is desirable to use analytical solutions for these disturbances, but these are only readily available for the inviscid, uniform-flow case.  Then, depending on the choice of base flow, an asymptotic error arises in that the base flow may not exactly approach uniform flow (e.g. if a boundary-layer solution is utilized), thus yielding ${\mathsfbi L} \boldsymbol{\hat q}^{\boldsymbol i} \ne 0$ in the far-field, indicative of an artificial source of scattered waves. For example, if a boundary-layer solution is used for the base flow, then there is a spurious source of $\mathcal{O}( \Rey^{-{1 \over 2}})$.  Moreover, our ${\mathsfbi L}$ includes viscous terms, and so ${\mathsfbi L} \boldsymbol{\hat q}^{\boldsymbol i} \rightarrow {1 \over \Rey}$ in the far-field rather than zero, and there are again artificial sources, which we term the viscous error. 

In the present work, we control both uniform-flow and viscous errors by choosing a sufficiently high Reynolds number such that the true sources are much larger than the spurious ones. More specifically, the choice of the Reynolds number restricts the maximum cross-stream wavenumbers of the ansatz assumed for $\boldsymbol{\hat q}^{\boldsymbol i}$ as explained next in \S~\ref{sec:opt_pw} and \S~\ref{sec:incident}. We verify this approach by comparing our solutions with ones where the region outside the boundary layer is artificially zeroed in \S~\ref{sec:in_out_fast}. In principle, there are more sophisticated ways of minimizing these errors, such as using a DNS for the base flow or by choosing incident waves that account for viscosity.

\subsection{Optimization}\label{sec:opt_pw}

We may write the incident wave as a sum of fundamental solutions to the (assumed inviscid) exterior (uniform flow) problem

\begin{equation}
\boldsymbol{\hat q}^{\boldsymbol i} = \sum^N_{j=1} a_j \boldsymbol{\psi_j} \equiv \tensor{\psi} {\boldsymbol a},
\end{equation}
where the $\boldsymbol{\psi_j}$ are each fundamental solution and are placed as columns of the matrix $\tensor{\psi}$. The specific form (plane waves) is enumerated in \S~\ref{sec:incident}. Now, let ${\mathsfbi B} \equiv - {\mathsfbi L} \tensor{\psi}$ so that \cref{eq:in)out} can be rewritten as
\begin{equation}
 {\mathsfbi L} \boldsymbol{\hat q}^{\boldsymbol s}  = {\mathsfbi B} {\boldsymbol a},
\end{equation}
where the vector of amplitudes ${\boldsymbol a}$ is the input to the linearized system (analogous to the input forcing fields in the unconstrained problem). 

We next define a global inner product
\begin{equation}
\langle \boldsymbol{b},\boldsymbol{d} \rangle = \boldsymbol{b^H} {\mathsfbi {W_{xyz}}} {\mathsfbi {W_e}}  \boldsymbol{d} = \boldsymbol{b^H} {\mathsfbi {W}} \boldsymbol{d}, 
\end{equation}
where $H$ is the Hermitian transpose and ${\mathsfbi {W}}$ is a positive-definite weight matrix. ${\mathsfbi {W}}$ is constructed as a product of ${\mathsfbi {W_{xyz}}}$, a diagonal positive-definite matrix of quadrature weights, and ${\mathsfbi {W_e}}$, an energy-weight matrix, so that $\langle \cdotp,\cdotp \rangle$ represents the volume-integrated quantity (up to a discretization error). The gain can thus be defined as a Rayleigh quotient
\begin{equation}\label{eq:gain1}
    G^2 =  \frac{\left< \boldsymbol{\hat q}^{\boldsymbol s}, \boldsymbol{\hat q}^{\boldsymbol s} \right>}{\boldsymbol{a^H} {\boldsymbol a}} = \frac{ {\boldsymbol{\hat q}^{\boldsymbol s}}^{\boldsymbol H} {\mathsfbi {W}} \boldsymbol{\hat q}^{\boldsymbol s} }{\boldsymbol{a^H} {\boldsymbol a}} = \frac{ \boldsymbol{a^H} {\mathsfbi {B^H}} {\mathsfbi {R^H}} {\mathsfbi {W}} {\mathsfbi R} {\mathsfbi B} {\boldsymbol a} }{\boldsymbol{a^H} {\boldsymbol a}},
\end{equation}
with optimal solution
\begin{equation}
   \{ \boldsymbol{\hat q}^{\boldsymbol {opt}}, {\boldsymbol a}^{\boldsymbol {opt}} \} = \mbox{argmax} \ G,
\end{equation}
where ${\mathsfbi R} = {\mathsfbi {L^{-1}}}$ is the global resolvent operator. In the optimization, we restrict $||{\boldsymbol a}||_2=1$ and scale each column of ${\mathsfbi B}$ so that $\left< \boldsymbol{b_j}, \boldsymbol{b_j} \right>=1$, which nullifies the arbitrary norm associated with $-{\mathsfbi L}\boldsymbol{\psi_j}$. Lastly, comparison to the unconstrained problem can be made by defining the following gains
\begin{equation}
 G^{c} = \left( { \left< \boldsymbol{\hat q}^{\boldsymbol s}, \boldsymbol{\hat q}^{\boldsymbol s} \right> / \  \left< \boldsymbol {\hat f}, \boldsymbol {\hat f}\right> } \right)^{1 \over 2} , \quad G^{uc} ={ \left< \boldsymbol{\hat q}^{\boldsymbol s}, \boldsymbol{\hat q}^{\boldsymbol s} \right>  }^{1 \over 2},
\end{equation}
where $\boldsymbol {\hat f} ={\mathsfbi B} {\boldsymbol a}$ for the constrained problem and where $\left< \boldsymbol{\hat f}, \boldsymbol{\hat f}\right>=1$ for the unconstrained optimization, thereby enforcing $G^c \le G^{uc}$.

To summarize, the scattered-wave ansatz allows us to constrain the optimization to realistic input forcings given by solutions to the outer problem in the form of plane acoustic, vortical, and entropic waves.  We will find linear combinations of such waves that maximize the amplification (according to the chosen norm) of the response.  The solutions can be directly compared with the worst-case disturbances for right-hand-side forcings that are not restricted to realizable disturbances to the outer problem.

\subsection{Incident waves}\label{sec:incident}

Plane acoustic waves in the uniform (assumed inviscid) free-stream take the form
\begin{equation}
    \boldsymbol {\psi^a}  = \boldsymbol{\hat q}^{\boldsymbol a} e^{\mathrm{i}(-\omega t +\alpha_a x+\kappa_a y + \beta_a z)}, 
\end{equation}
where $\alpha_a,\kappa_a,\beta_a \in\mathbb{R}$ are the acoustic wavenumbers in the $x$, $y$, and $z$ directions, respectively, and where  $\omega'^2 = c_\infty^2 \left( \alpha_a^2 + \kappa_a^2 +\beta_a^2 \right)$ and the amplitude $\boldsymbol{\hat q}^{\boldsymbol a}=\begin{bmatrix} 1 & {c_\infty {\alpha_a} \over \omega'} & { c_\infty {\kappa_a} \over \omega'} &  { c_\infty {\beta_a} \over \omega'} & (\gamma-1)T_\infty  \end{bmatrix}^T$, both with $\omega' = \omega-\alpha_a U_\infty$.  These waves satisfy the Euler equations linearized about a uniform flow (taken in the $x$-direction with speed $U_\infty$).
%\begin{equation}\label{eq:dispersion}
%    \omega'^2 = c_\infty^2 \left( \alpha_a^2 + \kappa_a^2 +\beta_a^2 \right), \quad \alpha_a,\kappa_a,\beta_a \in\mathbb{R},
%\end{equation}
%where $\alpha_a$, $\kappa_a$, and $\beta_a$ are the acoustic wavenumbers in the $x$, $y$, and $z$ directions, respectively. 

In the 2D case considered here, $\beta_a=0$, and the waves are parameterized with $\alpha_a$ (or a wave angle) at a specified real frequency, $\omega$.  The ranges of permitted values of $\alpha_a$ are based on the aforementioned dispersion relation for the different Mach-number regimes. For those cases where $|\alpha_a|$ is unbounded, we limit it to the highest wavenumber that can be resolved over 10 grid points, so that we take $|\alpha_a| \le \frac{2 \upi}{10 \Delta x} $.
%\begin{table}
% \begin{center}
%\def~{\hphantom{0}}
% \begin{tabular}{c|c}
%      $M_\infty$ & Range \\
%      \hline
%      $< 1$ & \( \displaystyle  -{1 \over 1 - M_\infty} \le {\alpha_a c_\infty \over \omega} \le {1 \over 1 + M_\infty} \) \\
%      $ = 1$ & \( \displaystyle  {\alpha_a c_\infty \over \omega} \le {1 \over 2}  \) \\
%      $ > 1$ & \( \displaystyle  {\alpha_a c_\infty \over \omega} \le {1 \over M_\infty+1 }  \) \& \( \displaystyle  {\alpha_a c_\infty \over \omega} \ge {1 \over M_\infty-1 }  \)
% \end{tabular}
% \caption{Wavenumber ranges for acoustic waves}
% \label{tab:acous_wavenum_tim}
% \end{center}
%\end{table}

Planar vortical and entropic wave solutions in the uniform free-stream are of the form
\begin{equation}
    \boldsymbol{\psi^{v,e}} = \boldsymbol{\hat q}^{\boldsymbol {v,e}} e^{\mathrm{i}(-\omega t +\alpha_{v,e} x+\kappa_{v,e} y + \beta_{v,e} z)},
\end{equation}   
where the amplitudes are $\boldsymbol{\hat q}^{\boldsymbol v}= \begin{bmatrix} 0 & - {\kappa_v + \beta_v \over \alpha_v} & 1 & 1 & 0 \end{bmatrix}^T$ and $\boldsymbol{\hat q}^{\boldsymbol e}= \begin{bmatrix} -1 & 0 & 0 & 0 & T_\infty \end{bmatrix}^T$, respectively. The wavenumbers $\alpha_{v,e}=\omega/M_\infty$, $\kappa_{v,e}$, and $\beta_{v,e}$ correspond to the Cartesian $x$, $y$, and $z$ directions, respectively, in which the latter two quantities are real but otherwise unconstrained. Realistic vortical and entropic disturbances will be compact and thus an infinite superposition of the plane waves.  However, decomposing the disturbances in Fourier modes has the advantage of identifying those wavelengths of disturbances to which the boundary layer is most receptive. As in the acoustic waves, we limit our attention to the 2D case ($\beta_{v,e}=0$) and set $\max(\kappa_{v,e})$ to the minimum of either those supported by at least 15 grid points within the boundary layer or satisfy $\Rey_{\lambda_{v,e}}\geq 2000$. The latter constraint is set to minimize the free-stream viscous error, while still retaining a broad spectrum for $\kappa_{v,e}$. 

\subsection{Computational details}

From now, we restrict our attention to strictly 2D, flat-plate boundary layers. The LNS equations are discretized with fourth-order central finite-difference schemes and closed with no-slip boundary conditions ($\hat{u}'=\hat{v}'=0$) and 1D inviscid Thompson characteristic boundary conditions \citep{Thompson_1987} at the wall-normal boundaries. Isothermal conditions ($\hat{T}'=0$) are enforced at the wall for the parametric study and those validating to \citet{Ma_Zhong_2005}, whereas adiabatic conditions ($\partial \hat{T}'/\partial y=0$) are used for all other analyses. We employ inlet and outlet sponges to model open boundaries. The full computational details of the code, CSTAT, are given in \cite{kamal_2021}.  

The computational domain contains wall-normal grid clustering in the boundary layer \citep{Malik_1990} and extends from $x^* \in [0.006,0.4]$ m and $y^* \in [0, 0.01]$ m with $N_x \times N_y = 3001 \times 250$. The base flow is computed using the Howarth–Dorodnitsyn transformation of the compressible Blasius equations. Note that each forcing vector $-{\mathsfbi L}\boldsymbol{\psi_j}$ is computed with a wall-normal resolution of $5N_y$ and interpolated back onto the stability grid to minimize free-stream discretization error. 

Different inner products (and associated norm) can be used to measure the strength of the response. Hereafter, we exclusively employ the Chu energy \citep{Chu1965} for both the forcing and response norms, which follows previous compressible input-output analyses of \citet{towne_2022,schmidt_2018,cook_nichols_2022}. The columns of the ${\mathsfbi B}$ matrix, which correspond to the scattered forcing fields, are therefore also normalized similarly. 

\subsection{Validation}\label{sec:validate}

We validate our methodology by comparing to DNS for a 2D Mach 4.5 adiabatic-wall, flat-plate boundary layer from \citet{Ma_Zhong_2003,Ma_Zhong_2003_P2,Ma_Zhong_2005}, which we subsequently refer to as MZ1, MZ2, and MZ3, respectively. A summary of the relevant computations from each paper is provided in \cref{tab:DNS_summary}. For validation purposes, we focus on the case where the boundary layer is excited by free-stream slow and fast acoustic waves at incident angles of $\theta_\infty^*=0^\circ$ and $\theta_\infty^*=22.5^\circ$, respectively, processed through an oblique shock using DNS. Although the shock is neglected in our computations, the linear theoretical formulation of \citet{mckenzie_1968} predicts the maximum deflection of fast acoustic waves with $\theta_\infty^* \in [0,90]^\circ$ to be just $\approx 1.24^\circ$. This is computed with a constant shock angle of $\theta_s^* \approx 13.69^\circ$ from figure 4 of MZ1. Furthermore, MZ2 found that for incident fast acoustic waves, the transmitted waves of the same type are responsible for synchronizing with the boundary-layer modes (explained further later), and thus the other wave-modes generated downstream of the shock are unimportant. Lastly, slow acoustic waves at $\theta_\infty^*=0^\circ$ impinging on the shock generates predominantly the same type of waves propagating nearly parallel to the wall (MZ3). We can thus neglect the shock in comparing our results to MZ2 and MZ3. 
\begin{table}
 \begin{center}
\def~{\hphantom{0}}
 \begin{tabular}{p{0.1\linewidth} | p{0.8\linewidth}}
      Paper & Relevant DNS computations \\
      \hline
      MZ1  & Steady-state base flow characterizing the oblique shock.  \\
      MZ2 & Wall-pressure response from an incident fast acoustic wave at $\theta_\infty^*=22.5^\circ$; quantification of the response of boundary-layer modes (Mode F1/F2 and second mode) to free-stream fast acoustic waves for $\theta^*_\infty \in [0,90]^\circ$.  \\
      MZ3 & Wall-pressure response from an incident slow acoustic wave at $\theta_\infty^*=0^\circ$. \\
 \end{tabular}
 \caption{Summary of relevant DNS computations performed by MZ1, MZ2, and MZ3 for a 2D Mach 4.5 adiabatic-wall, flat-plate boundary layer.}
 \label{tab:DNS_summary}
 \end{center}
\end{table}

In our computations, we force the LNS equations with $-{\mathsfbi L}\boldsymbol{\psi_j}$ corresponding to fast and slow acoustic waves at the aforementioned incident angles and compare the total solution ${\boldsymbol{\hat q}} = \boldsymbol{\hat q}^{\boldsymbol i} + \boldsymbol{\hat q}^{\boldsymbol s}$ to the DNS calculations. In comparing results, we adopt the following nomenclature from local linear stability theory (LST): Modes F1 and F2 are the sequential discrete modes emanating from the fast acoustic branch, whereas Mode S originates from the slow continuous spectrum, such that the second mode corresponds to Mode S during and post-synchronization with Mode F1. 
\begin{figure}
\centering
  \begin{subfigure}{0.425\textwidth} %0.4491
     \centering
    {\phantomsubcaption\label{fig:pwall1}
    \scalebarimg{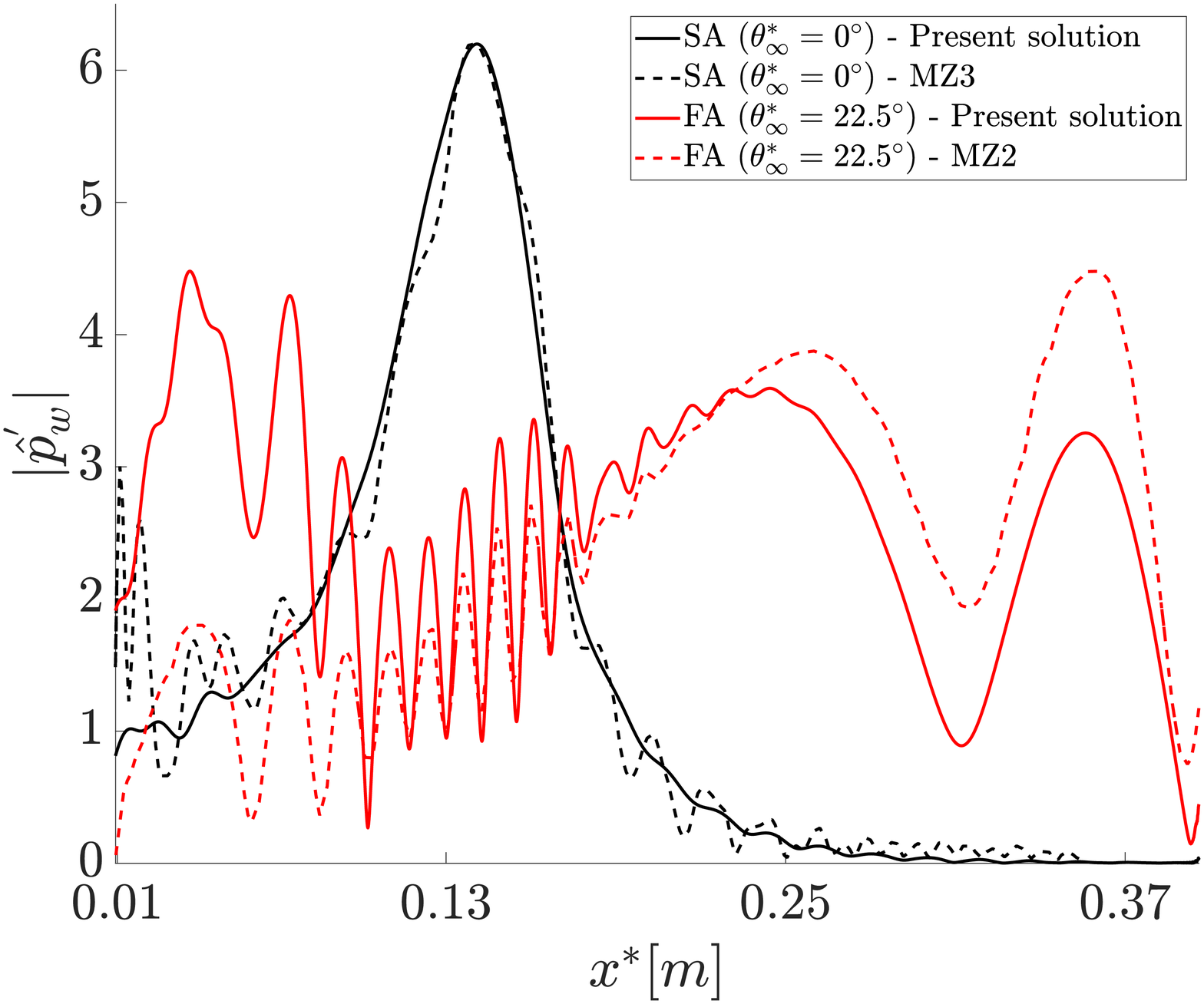}{1}{a}}
  \end{subfigure}
  \begin{subfigure}{0.45\textwidth}
     \centering
     \vspace*{-0.46cm}
     {\phantomsubcaption\label{fig:rho_resp_valid}
    \scalebarimgtr{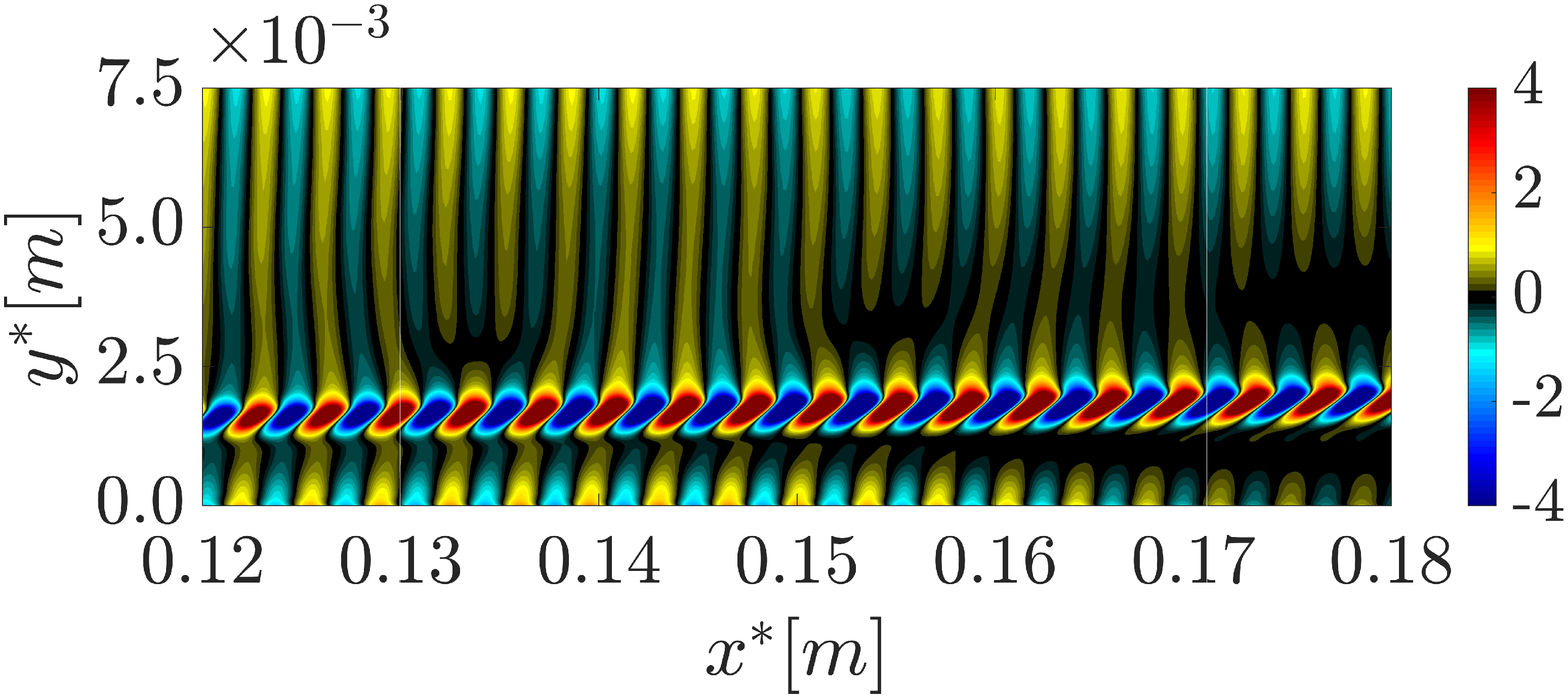}{1}{b}}
    %\vspace*{-0.05cm}
    \hspace*{1.4mm}
    \includegraphics[width=0.995\textwidth]{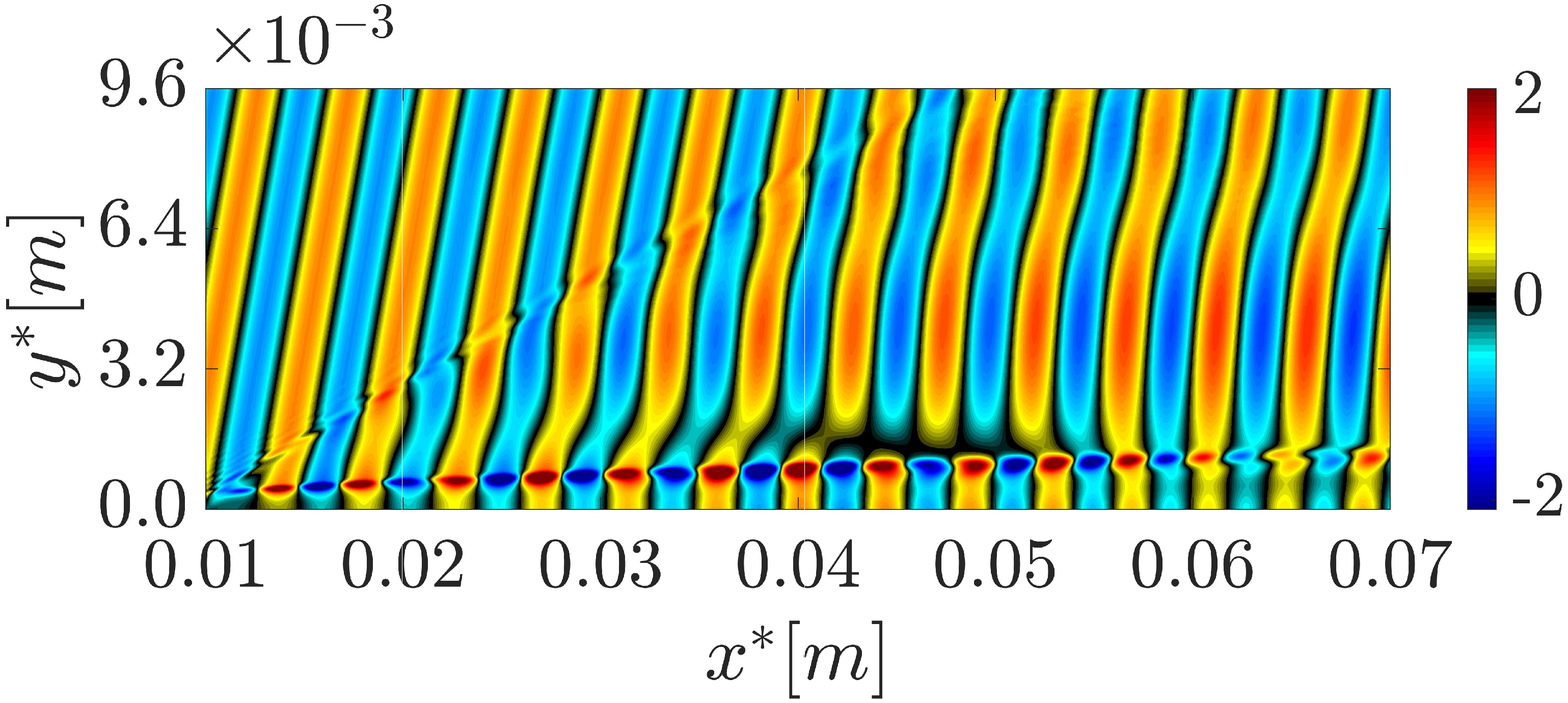}
  \end{subfigure}
  \vspace{-2.2\baselineskip}
  \caption{(\textit{a}) Wall-pressure amplitudes for the present ${\hat{\bf q}}$ solution compared to those of MZ2 and MZ3 with free-stream slow (SA) and fast (FA) acoustic waves at $M_\infty=4.5$ and $F=2.2 \times 10^{-4}$; (\textit{b}) the corresponding density responses for SA (top) and FA (bottom).}
  \label{fig:pwall}
\end{figure}
\Cref{fig:pwall}(\textit{a}) compares the wall-pressure amplitudes between DNS and the present solution at $F=\frac{\omega^*\nu^*_\infty}{{U_\infty^{*2}}}=2.2 \times 10^{-4}$. The pressure amplitude has been normalized to agree at the peak of each curve since both sets of computations are linear. For the DNS computations, linearity implies the non-dimensional amplitudes of the disturbances were at least one order of magnitude larger than the maximum numerical noise while also sufficiently small to remain in the linear regime (MZ1). Great agreement is observed for the slow acoustic wave and the agreement is satisfactory for the fast acoustic wave, especially in the region $0.1 < x^* < 0.2$ m, which corresponds to the location where the second Mack mode is dominant. We speculate the discrepancy in the leading-edge region is due to the shock in the DNS being locally oriented at $\theta_s^*  \approx 15.8^\circ$, which contrasts the global shock angle of $\theta_s^*  \approx 13.69^\circ$ used to estimate the maximum deflection of incident fast acoustic waves, resulting in larger local refraction when compared to further downstream. This likely effects the resonance with Mode F1 (the dominant mode near the inlet) since it exhibits higher sensitivity to incident-disturbance angles compared to the second mode (see \cref{fig:fast_acous}(\textit{a})). Finally, the density response for slow acoustic waves at $\theta_\infty^*=0^\circ$ in \cref{fig:pwall}(\textit{b}) matches well with the corresponding figure~11 of MZ3.

\section{Optimal global receptivity analysis}\label{sec:in_out_fast}

We now investigate the inverse problem of determining the linear combination of free-stream disturbances that lead to the maximal flow response.  We use the same base flow and parameters from \S~\ref{sec:validate}.

\begin{figure}
  \centering
  \begin{subfigure}[H]{0.3629\textwidth}
     \centering
    {\phantomsubcaption\label{fig:fast_acous2}
    \scalebarimg{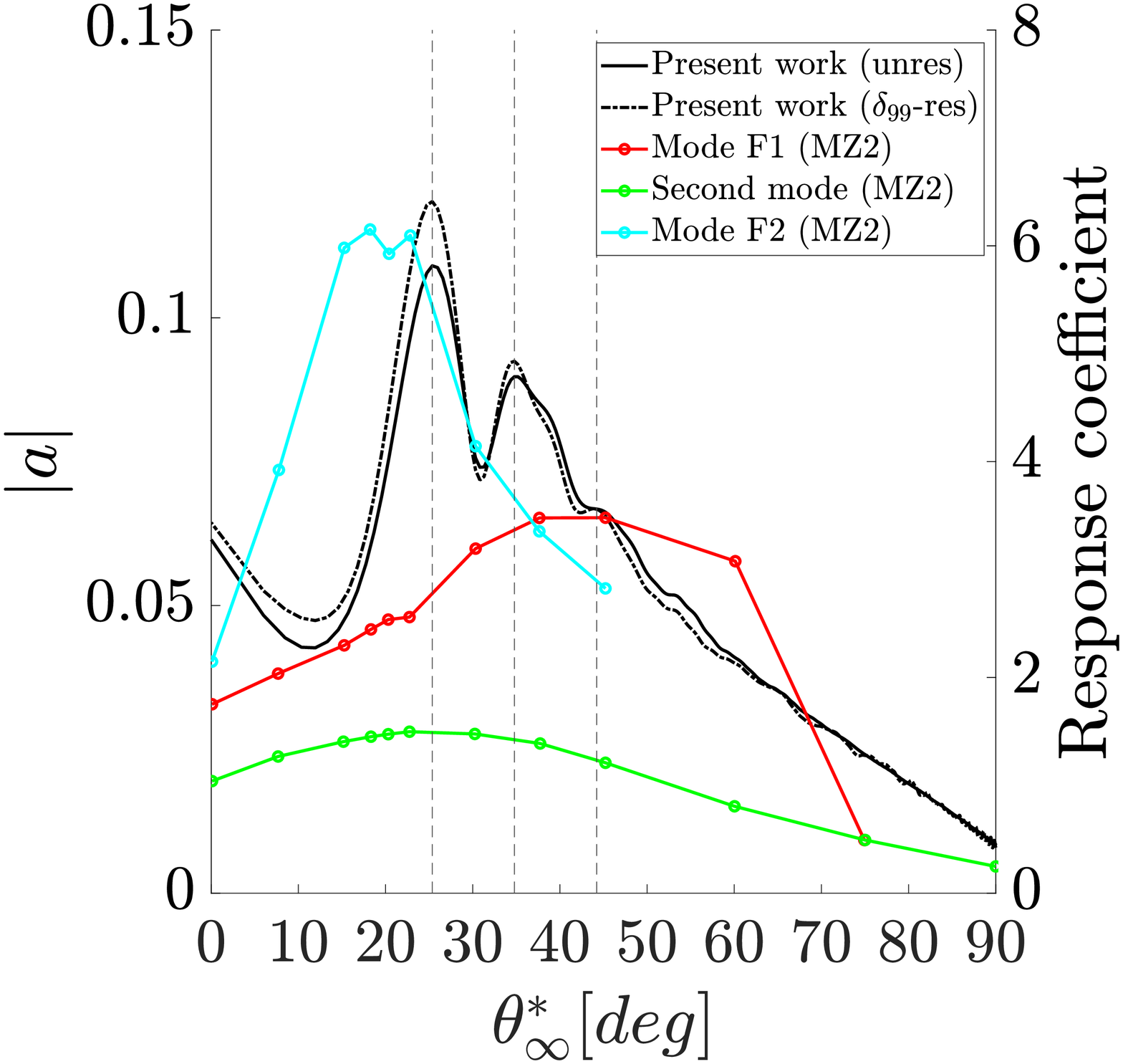}{1}{a}}
  \end{subfigure}
  \begin{subfigure}[H]{0.45\textwidth}
     \centering
      \vspace*{-0.45cm}
    {\phantomsubcaption\label{fig:fastacous_resp}
    \scalebarimgtr{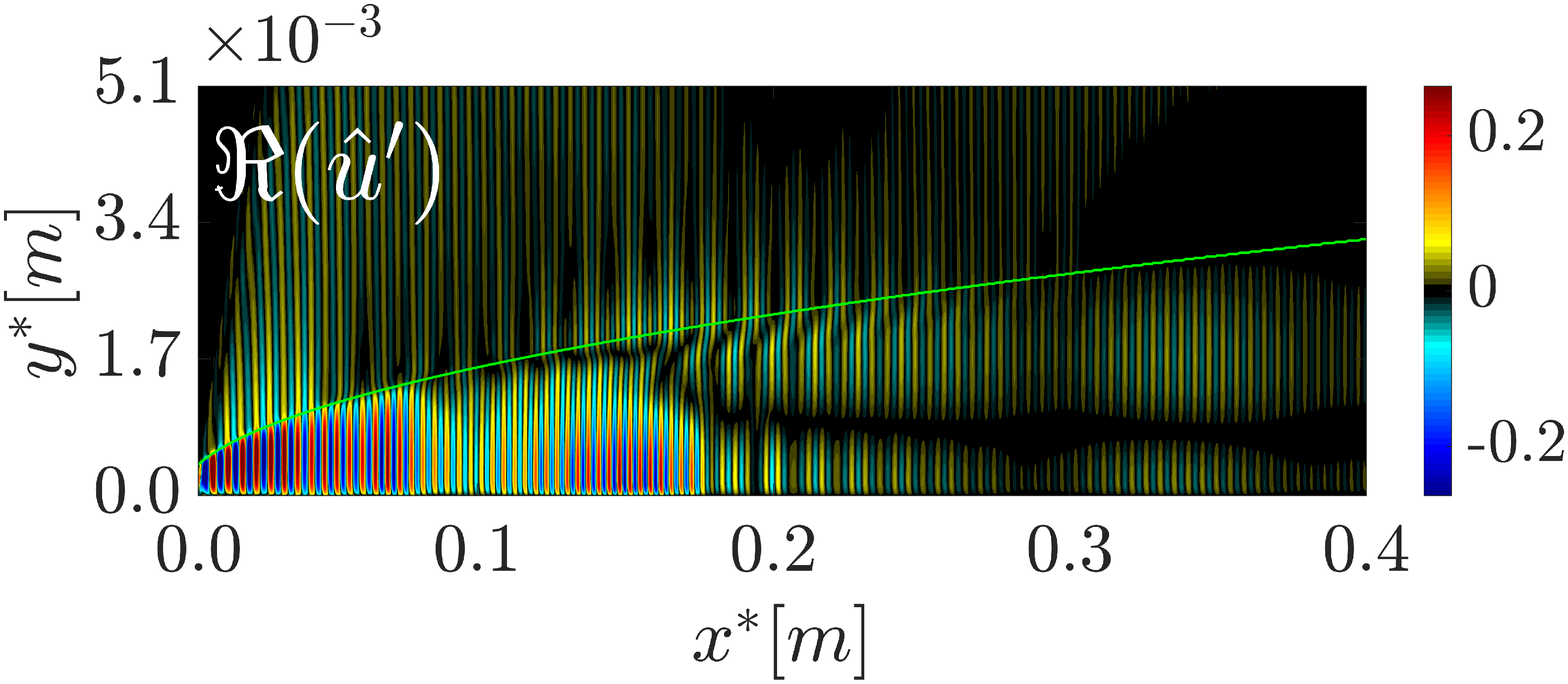}{1}{b}}
    \vspace*{-0.05cm}
    \hspace*{1.4mm}
   \includegraphics[width=0.995\textwidth]{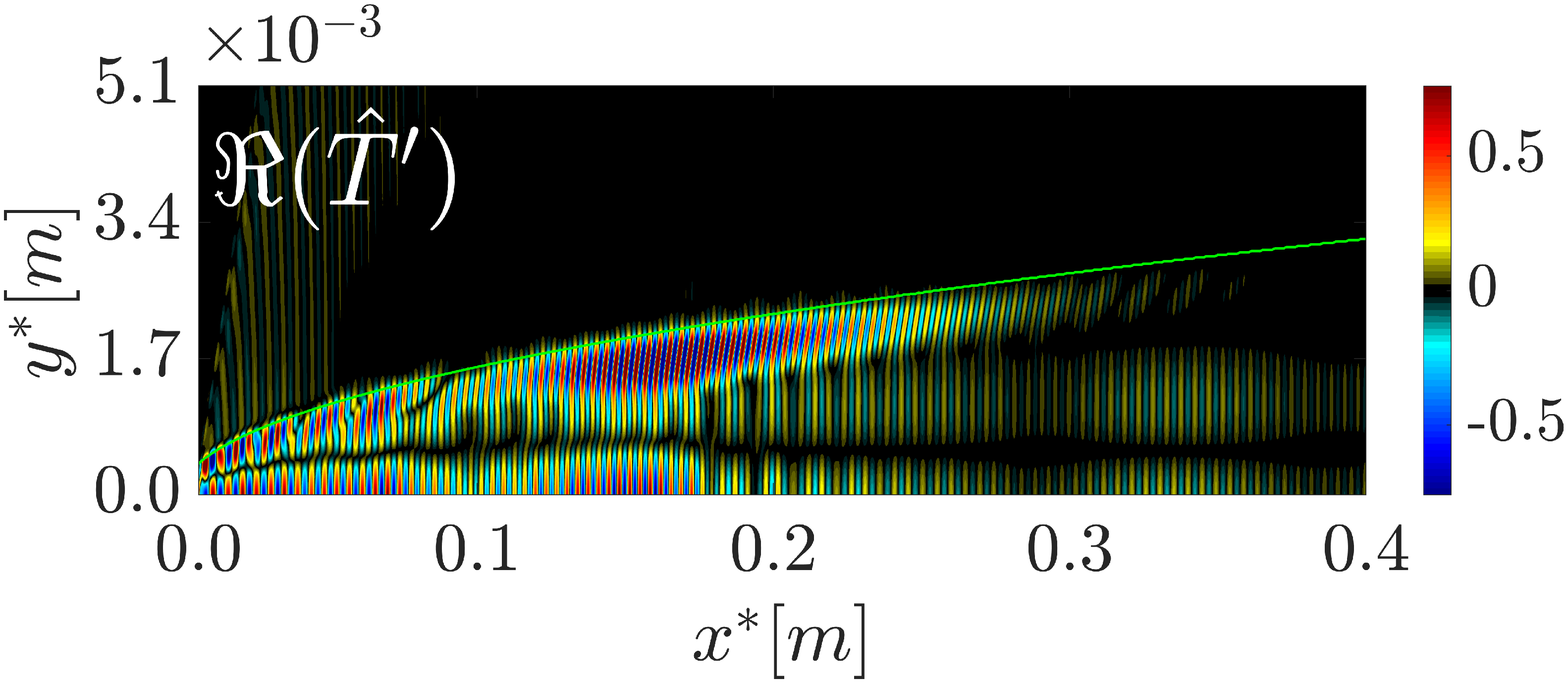}
  \end{subfigure}
  \vspace{-2\baselineskip}
  \caption{(\textit{a}) Optimal amplitude profile with free-stream fast acoustic waves at $M_\infty=4.5$ and $F=2.2 \times 10^{-4}$; (\textit{b}) the corresponding $\boldsymbol{\hat q}^{\boldsymbol s}$ responses (green isocontour is $\delta_{99}$). Colored lines in (\textit{a}) are the response coefficients from MZ2, the dashed lines are along the optimal angles from the scattering framework, and the dash-dotted line is the optimal amplitude profile with scattering sources restricted to $\delta_{99}$.}
  \label{fig:fast_acous}
\end{figure}

To allow comparison with the results of MZ2 for the forward problem, we initially restrict our attention to downstream-propagating fast acoustic waves from above the plate, i.e. $\alpha_a \geq 0$ and $\kappa_a \leq 0$, with $F=2.2 \times 10^{-4}$. We discretize the corresponding incident wave angles $0 \le \theta_\infty^* \le 90^\circ$ using $N = 1000$ points. The corresponding distribution of waves with amplitude $a$ is plotted against $\theta_\infty^*$ in \cref{fig:fast_acous}(\textit{a}) with prominent peaks observed at incident wave angles of $25^\circ$ and $35^\circ$, and a less significant peak at  $44^\circ$. We compare this curve to response coefficients computed by MZ2 for the forward problem computed over the range of angles. They measured approximate response coefficients for Modes F1 and F2 by using the maximum wave amplitudes in their respective dominant regions, according to LST, whereas the second-mode response coefficients were calculated by Fourier-transforming the pressure from the global DNS (at the specified frequency) and selecting the second-mode amplitude with its wavelength again inferred from LST.  

The comparison allows us to interpret the optimal solution as one that directly excites the second mode by selecting the fast acoustic waves at $\theta_\infty^*=25^\circ$, but also one that excites Mode F1 over a range of angles where its response coefficient is largest (and larger than the second mode). The higher response coefficient for Mode F1 is due to the synchronization between its wavenumber/wave speed and those of the free-stream fast acoustic waves at these angles, and largest near the leading edge due to the strongest base-flow non-parallelism, as is evident in \cref{fig:fast_acous}(\textit{b}). Downstream of the leading edge, the phase speeds of Modes F1 and S approach one another, and by $x^*\approx 0.11$ m, these two modes fully synchronize, which incites the second mode. The second mode remains unstable until it passes through the Branch II neutral point at $x^* \approx 0.155$ m (MZ2) and decays thereafter as Modes F1 and S de-synchronize, the latter of which is now the second mode. For $x^*>0.3$ m, the small growth and subsequent decay in \cref{fig:fast_acous}(\textit{b}) is due to the emergence of Mode F2 caused by the wavenumber/wave speed synchronization with the fast acoustic waves.

This importance of Mode F1 to second-mode amplification corroborates the finding of MZ2, and is further highlighted by comparing the respective gains from this optimal linear combination of fast acoustic waves, $G^c \approx 40$, with the gain obtained by limiting the input to only fast acoustic waves at $\theta_\infty^*= 25^\circ$, which we computed as $G^c \approx 21$, a reduction of about 48\%.

Lastly, we demonstrate how the true scattering sources in our framework are significantly larger than the spurious ones induced by the three sources of error mentioned in \S~\ref{sec:scattering} by repeating the above computation and artificially removing any sources outside the boundary layer. The corresponding amplitude profile is shown in \cref{fig:fast_acous}(\textit{a}) which is quantitatively similar to the original solution with $G^c$ only decreasing by $\approx 2\%$. 

\begin{figure}
  \centering
  \begin{subfigure}[H]{0.3629\textwidth}
     \centering
     {\phantomsubcaption\label{fig:kappa_evol}
    \scalebarimg{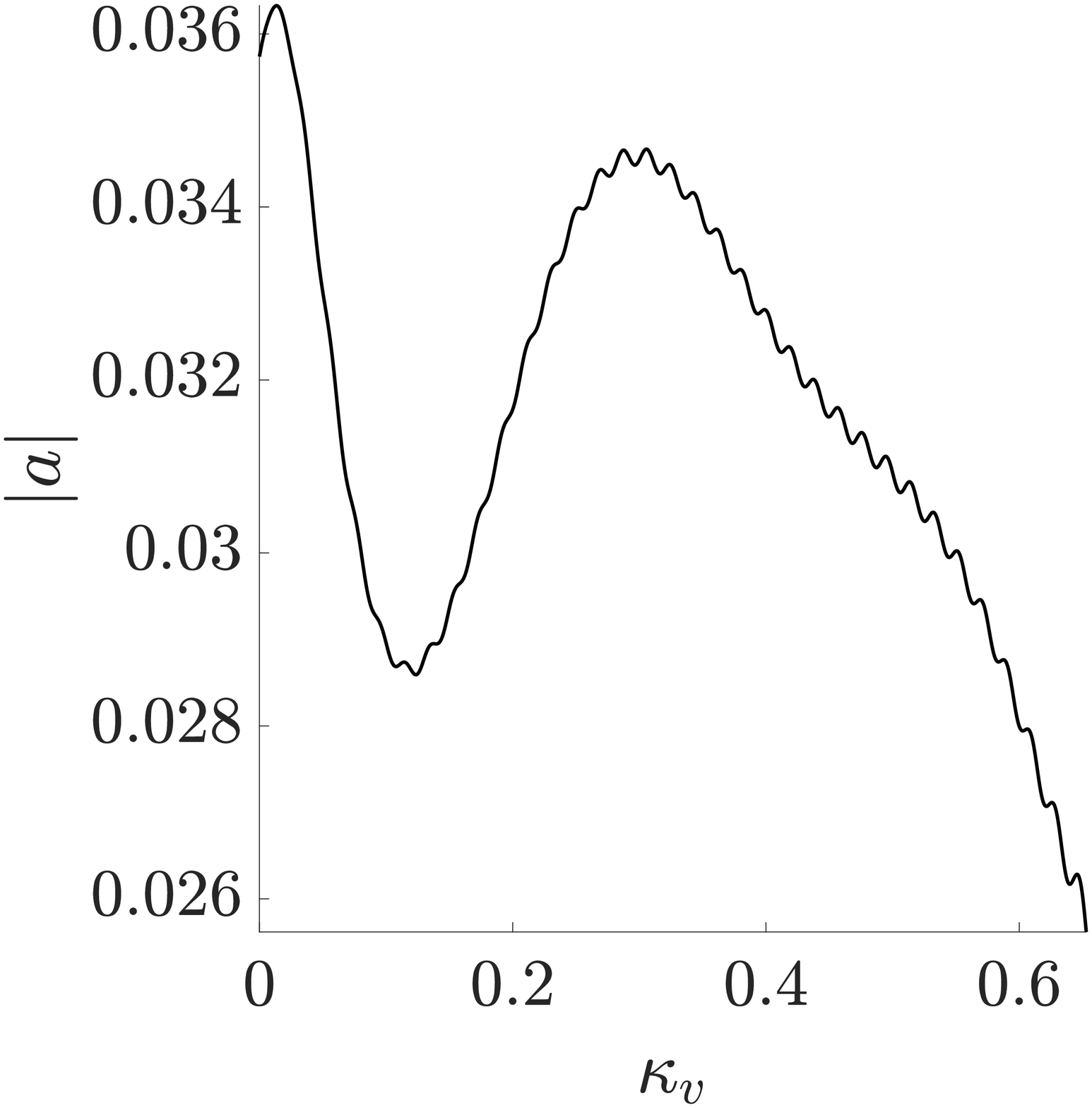}{1}{a}}
  \end{subfigure}
  \begin{subfigure}[H]{0.45\textwidth}
     \centering
     \vspace*{-0.455cm} %\vspace*{-0.475cm}
     {\phantomsubcaption\label{fig:vortentr_resp}
    \scalebarimgtr{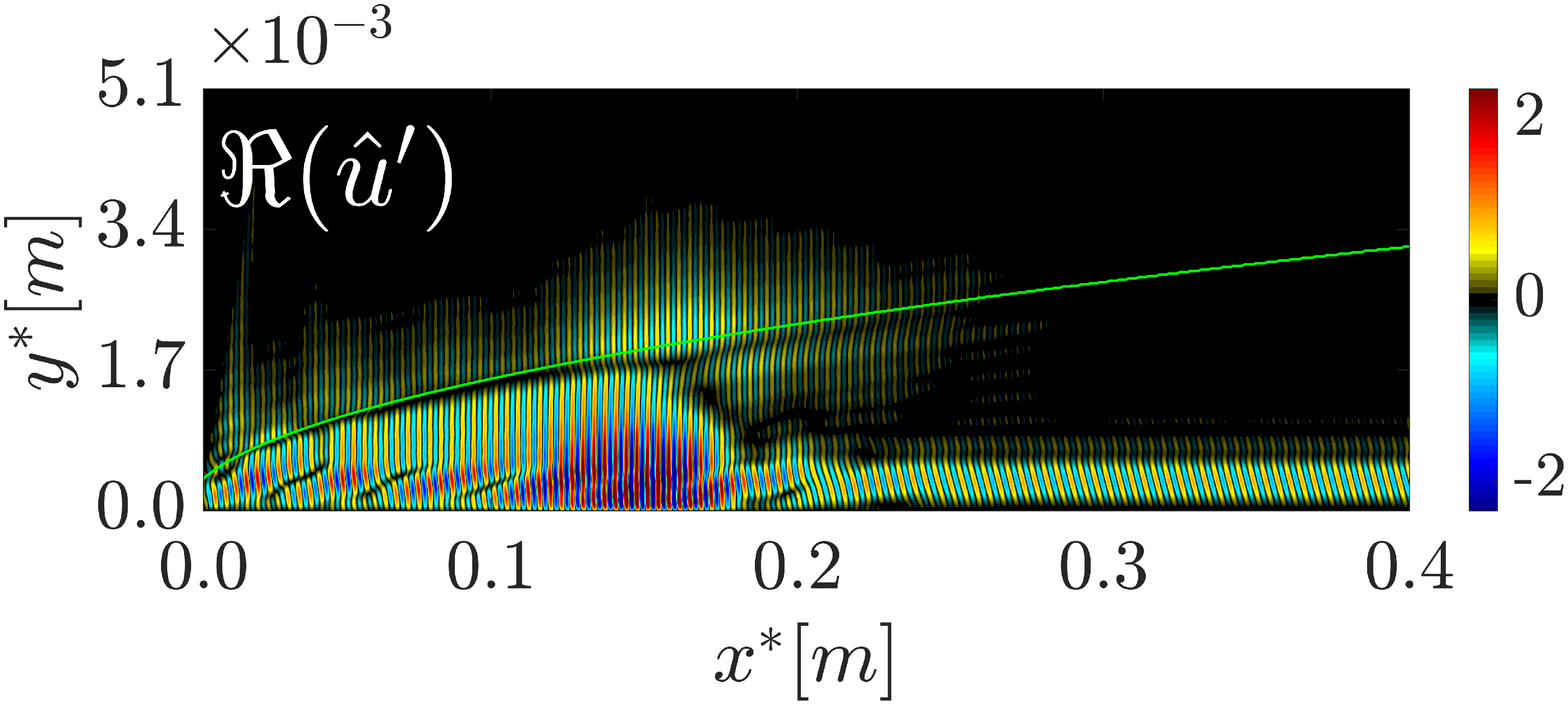}{1}{b}}
    \vspace*{0.15cm} %\vspace*{-0.1cm}
    \hspace*{1.4mm}
    \includegraphics[width=0.995\textwidth]{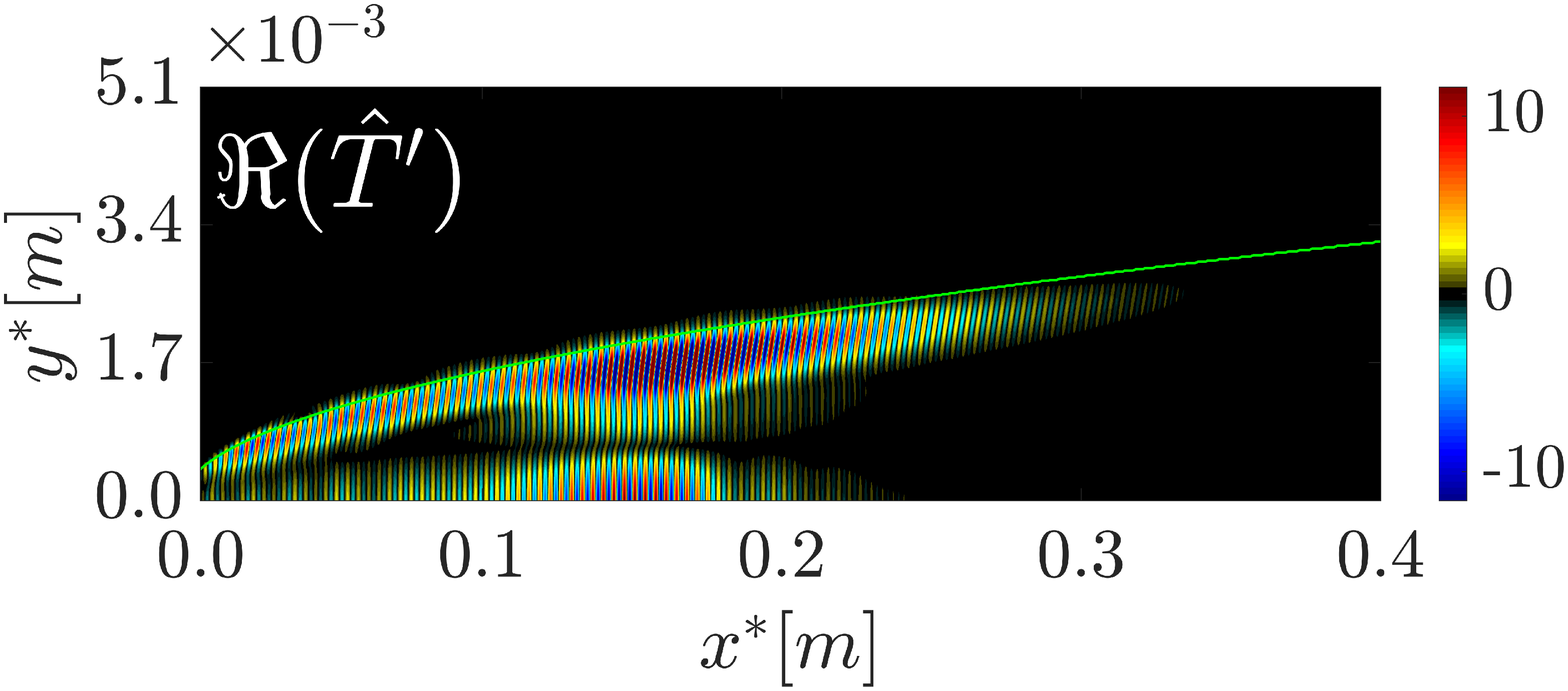}
  \end{subfigure}
  \vspace{-2\baselineskip}
  \caption{(\textit{a}) Optimal amplitude profile with free-stream vortical waves at $M_\infty=4.5$ and $F=2.2 \times 10^{-4}$; (\textit{b}) the corresponding $\boldsymbol{\hat q}^{\boldsymbol s}$ responses (green isocontour is $\delta_{99}$).}
  \label{fig:vortentr}
\end{figure}

We next analyse the case where the free-stream is restricted to vortical waves, again at $F=2.2 \times 10^{-4}$.  As investigated by \citet{schrader_2009}, there are two competing mechanisms for optimally perturbing the boundary layer using free-stream vortical disturbances: smaller wavelengths (large $\kappa_v$) are able to {\it penetrate} deeper into the boundary layer, but suffer faster viscous decay, whereas the opposite is true for larger wavelengths. The optimal distribution of vortical waves, shown in \cref{fig:vortentr}(\textit{a}), shows two maxima corresponding to $\kappa_v \approx 0.014$ and $\kappa_v\approx0.31$. Maximal excitation of disturbances is achieved by  simultaneously subjecting the boundary layer to highly-penetrating free-stream vortical modes and those that exhibit minimal viscous decay. Near the leading edge, free-stream vorticity penetrates the boundary layer and elicits a non-modal response characterized by large-scale streamwise jets emanating from the wall in the $\hat{u}'$ response field of \cref{fig:vortentr}(\textit{b}). These jets are also seen to be modulated by Modes F1 and S.  

Downstream of the leading edge, the phase speed of Mode F1 decreases, and by $x^*\approx 0.11$ m, Mode F1 synchronizes with Mode S to incite the second mode. During the second-mode growth however, the streamwise jets remain as seen in \cref{fig:vortentr}(\textit{b}). Once the second mode has decayed appreciably by $x^*\approx 0.18$ m, the jets are once again visible, but only weakly and for a short length as they suffer viscous decay. This is because free-stream vortical disturbances with $\kappa_v\approx 0.31$, which corresponds to $\lambda_v \approx 1.5\delta_{99}$ at the inlet, optimally excite the jets, but also experience relatively large viscous decay. 

Similar calculations were performed for slow acoustic waves, entropic waves, and for the gamut of all free-stream waves.  The respective gains are summarized in \cref{tab:gain_table}. Firstly, the slow acoustic waves yielded a gain $\approx 1.9$ times greater than the fast acoustic waves, which corroborates the general understanding that slow waves dominate acoustically-induced transition onset in adiabatic-wall high-speed boundary layers (MZ3, \citet{balakumar_2015}).  Vortical waves yielded a gain nearly identical to $G_{all}^c$, suggesting that the transient streamwise jets excited by vortical disturbances is the dominant receptivity mechanism for the current configuration. Entropy waves, on the other hand, play no significant role at these conditions. Lastly, the unconstrained optimization (standard input-output) problem yielded a gain nearly 300 times greater than $G_{all}^c$, suggesting that the physically-realizable inputs have a small projection onto the unconstrained (non-physically-realizable) inputs. Conversely, a large number of unconstrained forcing modes would be required to represent the physical forcing.  
\begin{table}
 \begin{center}
\def~{\hphantom{0}}
 \begin{tabular}{c|c}
      Case & $G^c$ or $G^{uc}$ \\
      \hline
      Fast acoustic ($\alpha_a \geq 0$ and $\kappa_a \leq 0$)  & 40 \\
      Slow acoustic ($\kappa_a \leq 0$) & 76 \\
      Vortical & 95 \\
      Entropic & 11 \\
      All & 96\\
     Unconstrained & $3.0 \times 10^{4}$
 \end{tabular}
 \caption{Gains from (un)constrained optimizations at $M_\infty=4.5$ and $F=2.2\times10^{-4}$.}
 \label{tab:gain_table}
 \end{center}
\end{table}

\begin{figure}
  \centering
    \includegraphics[width=0.4\textwidth]{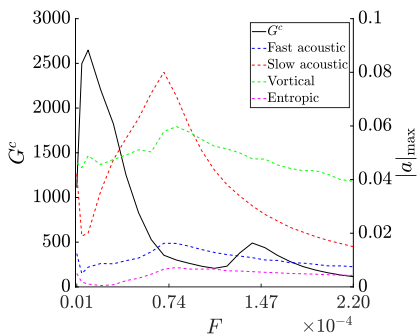}
  \vspace{-0.75\baselineskip}
  \caption{$G^c$ vs $F$ at $M_\infty=4.5$ with gamut of free-stream waves with the corresponding maximum amplitude from each wave-mode (fast/slow acoustic, vortical, and entropic).}
  \label{fig:gain_param}
\end{figure}

We now compute $G^c$ for the gamut of free-stream disturbances across a range of frequencies. Here, each wave-mode (fast/slow acoustic, vortical, and entropic) is discretized with $N=1000$ points, but in contrast to \S~\ref{sec:validate}, half the acoustic waves radiate above ($\kappa_a\leq0$) while the other half radiate below ($\kappa_a\geq0$) the plate. Additionally, $\alpha_a<0$ is included in the ansatz for fast acoustic waves. The gain profile and the maximum amplitude from each wave-mode, which reveals the dominant free-stream disturbance type, are shown in \cref{fig:gain_param}. The first mode is optimally excited at $F_{opt}\approx0.1\times 10^{-4}$, but the gain rapidly decreases with increasing frequency up until $F\approx0.7\times10^{-4}$. Then for $F > 1.1 \times 10^{-4}$, the second mode becomes the dominant instability and peaks at $F_{opt}\approx 1.4\times 10^{-4}$.

Across all frequencies, $|a|_\text{max}$ for entropic waves is the lowest, closely followed by fast acoustic waves. Although $|a|_\text{max}$ is attributed to a vortical wave at the lowest frequencies, the relative importance of slow acoustic waves steeply increases from $F \approx 0.05 \times 10^{-4}$ and eventually overtakes vortical waves at $F \approx 0.3 \times10^{-4}$, before becoming subdominant again by $F \approx 0.9\times10^{-4}$. The dominance of the vortical wave at high frequencies is likely attributed to the ``swallowing" effect \citep{fedorov_khokhlov_2001} where Mode F1 synchronizes with the continuous vorticity branch. The resulting Mode F1 waves, as discussed above, maximize the amplification of the second mode. While the fast acoustic waves are closer in wavenumber to Mode F1, the vortical waves are apparently more effective because they can simultaneously excite Mode F1 (``swallowing" effect) and Mode S (similar wavenumber).

\section{Discussion and future work}\label{sec:discussion}

We have developed a scattering ansatz to study global, optimal natural boundary-layer receptivity.  The technique can be understood as a generalization of receptivity theory to determine the linear combinations of free-stream disturbances that give rise to the maximal disturbance amplification in the boundary layer.  At the same time, it can be considered as a restriction of the forcing field in input-output analysis to a subspace associated with excitation by free-stream disturbances. As compared to many receptivity studies based on local methods, the global approach circumvents the need for asymptotic expansions. 

As a first application of the approach, we consider 2D disturbances to a Mach 4.5 flat-plate boundary layer for which the forward receptivity problem was previously solved using DNS (MZ2, MZ3).  The results validate the approach and revealed optimal disturbance amplification scenarios. When the free-stream is restricted to fast acoustic waves, maximal response is achieved by subjecting the boundary layer to acoustic waves with incident angles that optimally excite the second mode, but also, to a lesser extent, Mode F1. The receptivity mechanism vastly changes, however, in the case of incident vortical waves, where an optimal combination of highly-penetrating and minimally-decaying incident waves produce a transient response characterized by large-scale streamwise jets emanating from the wall and modulated by Modes F1 and S in the $\hat{u}'$ response field. Finally, the efficiency of our approach was demonstrated by computing the optimal receptivity for the same Mach 4.5 flat-plate boundary layer over a range of frequencies, highlighting where first- (low-frequency) and second-mode (high-frequency) instabilities are most receptive to different types of free-stream disturbances. While the configurations considered here are restricted, the methodology can be readily applied to 3D disturbances and more complex geometries, and the scattering ansatz can include sources associated with the shock and shock layer.

\section{Acknowledgements}
This work has been supported by The Boeing Company (CT-BA-GTA-1), ONR (N00014-21-1-2158), and NSERC (PGSD3-532522-2019).

\bibliographystyle{jfm}
\bibliography{jfm-instructions}

\end{document}